# Vanadium in yttrium aluminum garnet: charge states and localization in the lattice

V. Laguta[1], M. Buryi[1], A. Beitlerova[1], O. Laguta[2], K. Nejezchleb[3], M. Nikl[1],

[1]Institute of Physics of the Czech Academy of Sciences, Cukrovarnická 10/112, 162 00 Prague, Czechia
[2]Institute of Physical Chemistry, University of Stuttgart, Pfaffenwaldring 55, D-70569 Stuttgart, Germany
[3]CRYTUR Ltd., Turnov, Czechia

**Abstract**

Vanadium ions charge states and their incorporation in the yttrium aluminum garnet $Y_3Al_5O_{12}$ (YAG) lattice were studied by the correlated optical absorption and electron paramagnetic resonance (EPR) measurements. In as-grown crystals, the occupation of the $V^{3+}$ at both the octahedral and tetrahedral aluminum sites was proven. The $V^{3+}$ to $V^{4+}$ charge transformation was observed after annealing in air, whereas annealing in the hydrogen atmosphere resulted exclusively in a slight weakening of the $V^{3+}$ absorption bands due to partial recharge of these ions. Spin Hamiltonian parameters of the $V^{3+}$ and $V^{4+}$ ions at the tetrahedral sites including the zero field splitting and the $^{51}V$ hyperfine constants have been determined using the high-frequency, up to 300 GHz, EPR measurements. From the analysis of the spin Hamiltonian parameters in the framework of the crystal field theory, the ground state energy levels splitting of the $V^{3+}$ and $V^{4+}$ ions were calculated. The charge distribution over the tetrahedral $V^{3+}$ and its nearest oxygen surroundings was found to be strongly inhomogeneous whereas the tetrahedral $V^{4+}$ ion concentrated the charge with very weak participation of surrounding ligands. Furthermore, the correlation of the optical and EPR data allowed the proper assignment of the optical absorption peaks in YAG:V crystals.

KEYWORDS: A. optical materials, B. optical properties, C. electron spin resonance, D. defects

## 1. Introduction

High power compact lasers constructed on the base of microchip geometry [1] remain demanding sources of optical irradiation in different branches of science and industry. The most frequent applications are micromanufacturing, remote sensing, data storage devices, etc. [2]. A simple scheme of microchip laser represents a diode pumped solid state laser passively Q switched by a solid state saturable absorber. One of the first materials used as a Q switch was $Y_2SiO_5$:Cr (YSO) crystal coupled to Cr:$LiCaAlF_6$ (LiCAF) laser [3]. Nowadays, the $Y_3Al_5O_{12}$:V (YAG) is widely used as a passive Q-switch, demonstrating wideband saturable absorption [4-8]. The effective passive Q switching and mode locking were demonstrated by YAG:V in 1.3 µm wavelength diode-pumped Nd:$YVO_4$ lasers [9-11] and lamp-pumped $YAlO_3$:Nd (YAP) lasers [12]. The 1.3 µm lasers are perspective due to their potential applications in e.g., fiber communication technique, medical treatment, chemical detection, scientific research [13-15]. Other examples of gain media employing the effective passive Q switching by means of YAG:V are Yb- or Nd-doped crystals such as Y(Lu)AG:Nd(Yb), YVO4:Nd and (KGd(WO$_4$)$_2$:Nd (KGW), GdVO$_4$:Nd [9, 16-20]. A special case of the YAG:Nd ceramic laser Q switched by the YAG:V operating at 1357 nm is reported relatively recently [21]. YAG:V also finds application as a saturable absorber in lasers operating at other than 1.3 µm wavelengths, e.g., 1.44 µm giant pulse generation in YAG:Nd [22] or 760–800 nm LiCAF:Cr lasers [23].

Despite the technical importance of the YAG:V crystals, there are only few studies (mainly by optical absorption spectroscopy) in the literature addressing the vanadium ions charge state and their incorporation in the garnet host [24-26]. YAG has a cubic crystal structure belonging to the space group $O_h(10) - Ia3d$ [27]. There are 24 tetrahedral (*d*) and 16 octahedral (*a*) sites per unit cell with $O^{2-}$ ions at the vertices and $Al^{3+}$ at the centers of these sites. Each octahedron is stretched along one of its threefold axes and has site symmetry $C_{3i}$ with the trigonal axes coinciding with one of the [111] directions of the crystal. The tetrahedra are also distorted by stretching along one of their fourfold axes and have site symmetry $S_4$, with the tetragonal axes along one of the [100] directions of the crystal. The axial distortions are fairly strong and give rise to an axial term in the crystal field for each Al site. Previous studies [24-26] have shown that the small vanadium ions can occupy both the aluminum tetrahedral and octahedral sites in YAG. The only tetrahedral $V^{3+}$ ($V_T^{3+}$) was tentatively suggested to be the source of the effective absorption in infrared region exhibiting three broad bands with maxima at approximately 820 nm, 1150 nm and 1300 nm [24].

Being a transition metal ion, vanadium can easily change its valence state with respect to local charges following the electroneutrality preservation. Therefore, the vanadium charge state transformation of a different kind in the YAG lattice is of great interest. In general, the actual structure of vanadium centers and their simple and effective monitoring is of particular importance in order to further improve saturable absorber parameters. Electron Paramagnetic Resonance (EPR) experiment supplemented by optical absorption measurements is the strongest tool to investigate this problem. Note that the optical absorption itself cannot unambiguously distinguish different valence states of vanadium and especially its different surroundings. In this respect, providing much better resolution of spectral lines from different $V^{3+}$ and $V^{4+}$ centers EPR can be extremely useful in the identification and monitoring of the vanadium centers. Nevertheless, to the best of our knowledge, only one work dedicated to the EPR study of the octahedral $V^{3+}$ ($V_O^{3+}$) accompanied by the absorption measurements was published so far [26]. Moreover, there is a very limited number of EPR identification of the $V^{3+}$ ions in non-cubic ligand surroundings because of large zero-field splitting in crystal field.

Consequently, the aim of this work is a detailed study of the $V^{3+}$ and $V^{4+}$ centers in YAG crystal, clarification of their ground state energy levels diagrams and $V^{3+}/V^{4+}$ transformation under various technological treatments by using correlated optical absorption and electron paramagnetic resonance measurements.

The paper is organized in the following way: After an introduction to the problem and a short description of the applied experimental methods (Sec. 2), we first report the optical absorption data on YAG:V crystals subjected to various technological treatments (Sec. 3.1). Then we present our EPR data on $V^{3+}$ and $V^{4+}$ centers (Sections 3.2–3.4). Finally, in Sec. 4, we draw conclusions.

## 2. Samples and experimental setup

Single crystals of vanadium doped yttrium aluminum garnets were grown by the Czochralski technique from 5N $Y_2O_3$, 5N $Al_2O_3$ and 4N $V_2O_5$ raw powders in a molybdenum crucible under reducing atmosphere by CRYTUR, Ltd. (Turnov, Czech Republic). The following samples were obtained: YAG1AG (as-grown), YAG1HA (post-growth annealed in $H_2$ atmosphere at elevated temperature), YAG2AA (post-growth annealed in air at elevated temperature). For measurements, the crystals were x-ray diffraction oriented, polished and cut in the (100) and (110) planes in a typical shape of about 2x2.5x6 $mm^3$. The same crystals were also used in the optical absorption measurements.

Optical absorption was measured by Shimadzu 3101PC UV spectrometer. EPR spectra were acquired in the X-Band (9.2-9.3 GHz) with the standard 3 cm wavelength EPR spectrometer in the temperature range 4.2 – 60 K using an Oxford Instruments liquid helium cryostat.

Additional EPR measurements were performed with the home-made high frequency spectrometer operated at frequencies from 82.5 GHz up to 1100 GHz in the magnetic field 0 − 15 T and temperatures 10 – 300 K. The field sweep resolution was 28000 points for the 0− 15 T region with the sweep time of 100-150 min for one spectrum. Details of this spectrometer design can be found in [28] and in supplementary material to this publication.

## 3. Results and discussion

### 3.1. Optical absorption spectra

Room temperature (RT) optical absorption spectra measured within 300-2000 nm in the YAG1AG (as-grown), YAG1HA (post-growth annealed in $H_2$ atmosphere), and YAG1AA (post-growth annealed in air) samples are shown in Fig. 1.

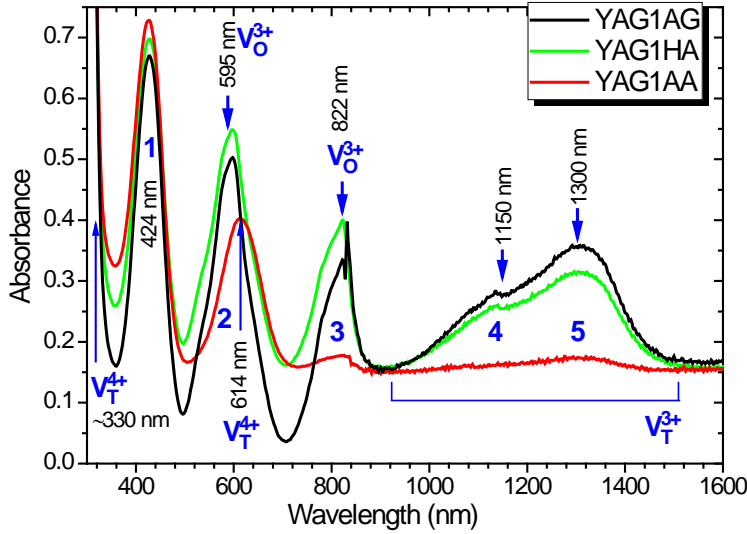

Fig. 1. RT absorption spectra of YAG:V crystals measured in as-grown (black dotted line), annealed in $H_2$ (green solid line), and annealed in air (red solid line). Bands corresponding to each charge state and localization of vanadium ions are indicated and numbered. Local maxima are stressed.

Five bands with maxima at about 424 nm (1), 598 nm (2), 822 nm (3), 1150 nm (4) and 1300 nm (5) can be clearly resolved in the spectra. There are no other transitions above 1600 nm. All bands seem to be complex, composed of several overlapping peaks. The strongest absorption is observed for the bands 4 and 5 in the as-grown YAG1AG and YAG1HA samples. Annealing in $H_2$ slightly reduces these bands amplitudes, however, the significant decrease is obtained after annealing in air along with large absorption drop also in the 822 nm ("3") and partially in the ~600 nm ("2") bands (it consists of at least two superimposed peaks at 595 nm and 614 nm) whereas the 424 nm ("1") peak survives unchanged both in the $H_2$ and air annealed samples. This can be explained by the bleaching of the vanadium (III) ions occupying the tetrahedral and octahedral aluminum sites, $V_T^{3+}$ and $V_O^{3+}$, respectively, by the annealing in air (Fig. 1), whereas the remained peaks should be ascribed to the vanadium (IV) ion in the tetrahedral surrounding, $V_T^{4+}$. In particular, the band 2 is red shifted by approximately 20 nm with respect to its initial spectral position and it becomes broader. Probably it is a new band compared to the as-grown and hydrogen annealed samples belonging namely to the $V_T^{4+}$. The roughly similar suggestion was made in [24]. According to the interpretation given in [24] and references therein for the vanadium ions in YAG structure the bands 1,2 peaking at ~425 nm and ~615 nm, respectively, were presumably attributed to the $V_O^{3+}$ previously basing on corundum analysis [24]. However, since the 595 nm peak in Fig. 1 is significantly suppressed after the air annealing, but the 424 nm one remains, they scarcely can be attributed to the same type of vanadium center taking into account the evidence of the $V_O^{3+}$ charge transformation under the annealing in air confirmed by EPR (Sec. 3.4). In particular, the $V_T^{3+}$ and $V_O^{3+}$ EPR spectra disappear after the annealing in air. The $V_T^{4+}$ spectrum appears instead so it does not exist in the as-grown and H$_2$ annealed samples at the marked concentration. Therefore, namely the 595 nm and 822 nm (not 614 nm) peaks were attributed to the $V_O^{3+}$ center, the very broad bands 4 and 5 were ascribed to the $V_T^{3+}$ while the 614 nm peak is assumed to be originated from the $V_T^{4+}$. For the tetrahedral vanadium (IV) ion two transitions are expected [24]. Since the 424 nm peak underwent no drastic changes after the annealing procedures it cannot be considered as the one belonging to the vanadium 4+ in the tetrahedral surrounding. This can be the $V^{4+}$ in octahedral environment responsible for it. More precise assignment of single peaks will be discussed also in more detail in the sections 3.3 - 3.4.

### 3.2. EPR of $V^{3+}$ octahedral center

EPR spectrum of the YAG1AG sample contains only the $V_O^{3+}$ and $V_T^{3+}$ (Figs. 2 and 3a)) resonances as well as the Mo$^{3+}$ lines (not visible in Fig. 2) studied previously [29-30] so they are not the subject of the present paper. Here the Mo$^{3+}$ is an uncontrolled impurity penetrated the material from molybdenum crucible. The $V_O^{3+}$ spectrum was identified previously [26]. As it will be shown below, the broad resonance line at the magnetic field of 850 mT belongs to $V_T^{3+}$.

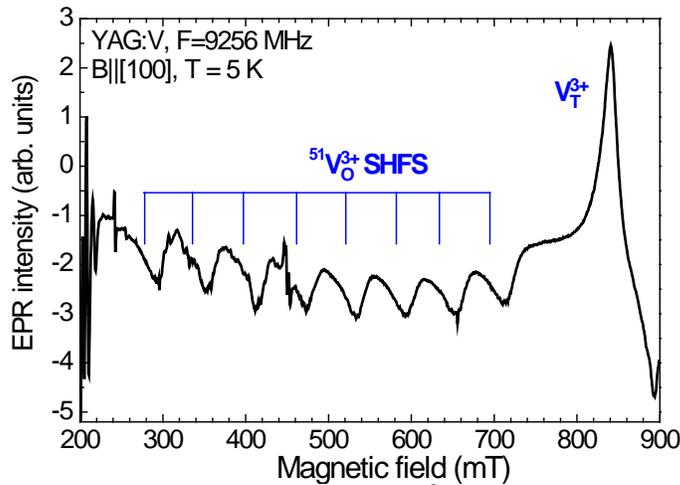

Fig. 2. Octahedral and tetrahedral V$^{3+}$ EPR spectrum in as-grown YAG crystal.

Each vanadium transition was recognized by the pronounced hyperfine (HF) structure counting eight lines of nearly equal intensity which originate from the interaction of the electron spin with the nuclear spin of $^{51}$V nucleus ($I = 7/2$ and 100% natural abundance). This property strongly confirms that the observed spectral lines indeed belong to vanadium ions.

The broad octahedral $V^{3+}$ spectrum measured at the **B** || [100] orientation of an external magnetic field clearly observed only at very low, T = 5-10 K, temperatures as shown in Fig. 2. At this chosen orientation, the $V^{3+}$ spectrum is almost undistorted as it does not overlap with other signals. At the temperature above 15 K, the $V_O^{3+}$ spectrum becomes very broad due to the short spin-lattice relaxation time. This agrees well with earlier studies of the $V^{3+}$ spectrum in corundum [31], where the $V^{3+}$ also occupies the six-fold coordinated Al site. The octahedral crystal field splits the $^3F$ free $V^{3+}$ ion ($3d^2$, $S = 1$) ground level into the three $^3T_{2g}$, $^3T_{1g}$ and $^3A_{2g}$ states (see, e.g. Fig. 7.6 of Ref. 32). The ground state is triply degenerated $^3T_{2g}$. Tetragonal or trigonal distortion partly removes the degeneration, so the new ground state is the $^3A_{2g}$ singlet. The spin-orbit coupling then results in the $S = 1$, represented by the $|0\rangle$, $|1\rangle$ and $|-1\rangle$ spin states [31,32] with the large, more than 7 cm$^{-1}$ zero-field splitting.

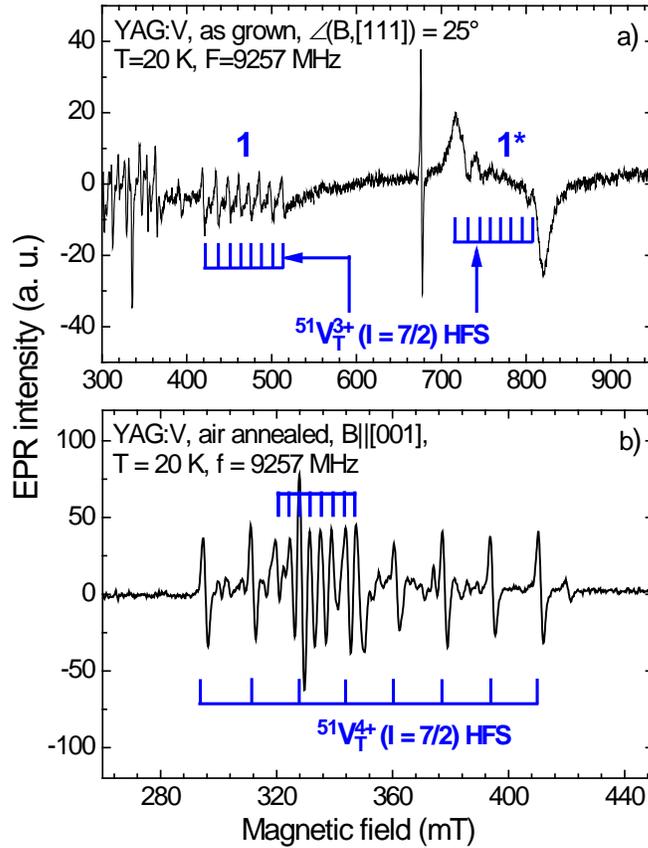

Fig. 3 EPR spectra of $V^{3+}$ and $V^{4+}$ in tetrahedral Al sites in YAG samples: a) as-grown and b) air annealed. 1 and 1* design spectral lines of EPR transitions of $V^{3+}$ ion.

Therefore, the observed $V_O^{3+}$ resonances (Fig. 2) at low microwave frequencies (X microwave band) occur only between the $|+1\rangle$ and $|-1\rangle$ states with the corresponding $\Delta M = 2$ spin eigenvalues difference [33]. The spin Hamiltonian parameters $g_{\parallel}^{eff} = 2.234$ and $A_{\parallel} = 0.033$ cm$^{-1}$ were derived by using the $h\nu = 2(g_{\parallel}\beta_e B + A_{\parallel}m)$ expression, where $\beta_e$, $g_{\parallel}$, $A_{\parallel}$ are the Bohr magneton, the g factor and hyperfine constant along the trigonal axis, $g_{\parallel}^{eff} = 2g_{\parallel}$. $B$ is a resonance magnetic field. The spin Hamiltonian parameters are in perfect agreement with the $g_{\parallel}^{eff} = 2.237$ and $A_{\parallel} = 0.033$ cm$^{-1}$ reported earlier for the octahedral V$^{3+}$ in YAG [26].

### 3.3. EPR of V$^{3+}$ tetrahedral center

Similarly to the $V_O^{3+}$ center, only one transition is detected in the EPR spectra attributed to the tetrahedral V$^{3+}$ center. Its spectral lines with the $^{51}$V HF structure are labelled as 1 and 1* in Fig. 3a for the chosen magnetic field direction, 25º from the [111] crystal direction in the (111) plane. They originate from three magnetically inequivalent positions of the $V_T^{3+}$ having mutually different orientations of the principal axes allowed for a tetrahedral aluminum site in the YAG lattice, namely along the <100> cubic directions.

Angular dependencies of the $V_T^{3+}$ resonance fields were measured in the $(1\bar{1}0)$ rotational plane since in this plane spectral lines from centers with different orientation of the principal axes are well distinguished. Their angular dependencies, corresponding to only electron spin transitions (i.e. center of gravity of HF structure), are shown in Fig. 4.

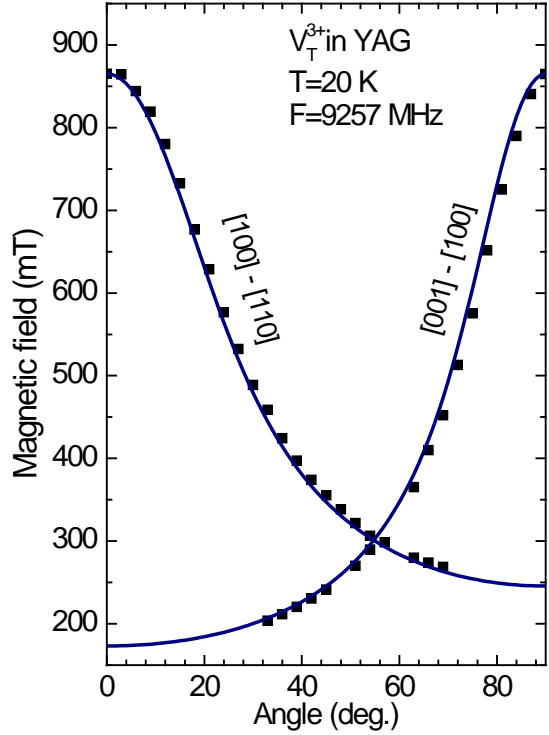

Fig. 4. Angular dependencies of the center of gravity of the $V_T^{3+}$ resonance lines measured in the $(1\bar{1}0)$ rotational plane (solid squares) and their fit to Eq. 1 (solid lines). In this rotation plane, magnetic field rotates from [001] to [100] direction for one center and from [100] to [110] direction for two other centers which resonances coincide.

In Fig. 4, one curve of the angular dependency corresponds to the rotation of the magnetic field from the z axial axis of one center (z-axis is along the [001] crystal direction) to the x-axis ([100] direction). The second curve corresponds to the rotation of the magnetic field from the x-axis of the other two centers (z-axis is along the [100] and [010] directions) to the [110] direction. The hyperfine structure has not been fitted as it is weakly resolved at most orientations of the magnetic field. The HF tensor will be determined from the measurements carried out at 300 GHz frequency.

Because at 9.257 GHz frequency only one forbidden transition $|+1\rangle \leftrightarrow |-1\rangle$ of the $V_T^{3+}$ center is observed, the spin Hamiltonian parameters of this center cannot be determined. Therefore, we undertook measurements at higher microwave frequencies, namely at 200 − 300 GHz. As an example, such a spectrum measured at 300 GHz at temperature 12-14 K and magnetic field direction along the [100] cubic axis is shown in Fig. 5. At this crystal orientation, one center has its axial z-axis parallel with magnetic field B, and two other centers have axial axes perpendicular to the magnetic field. Altogether, there are six groups of spectral lines with HF structure corresponding to two allowed $|\pm 1\rangle \leftrightarrow |0\rangle$ transitions and one forbidden $|-1\rangle \leftrightarrow |+1\rangle$ transition. The assigning of these transitions to measured spectral lines is shown in the bottom spectrum in Fig. 5, which presents the calculated spectrum.

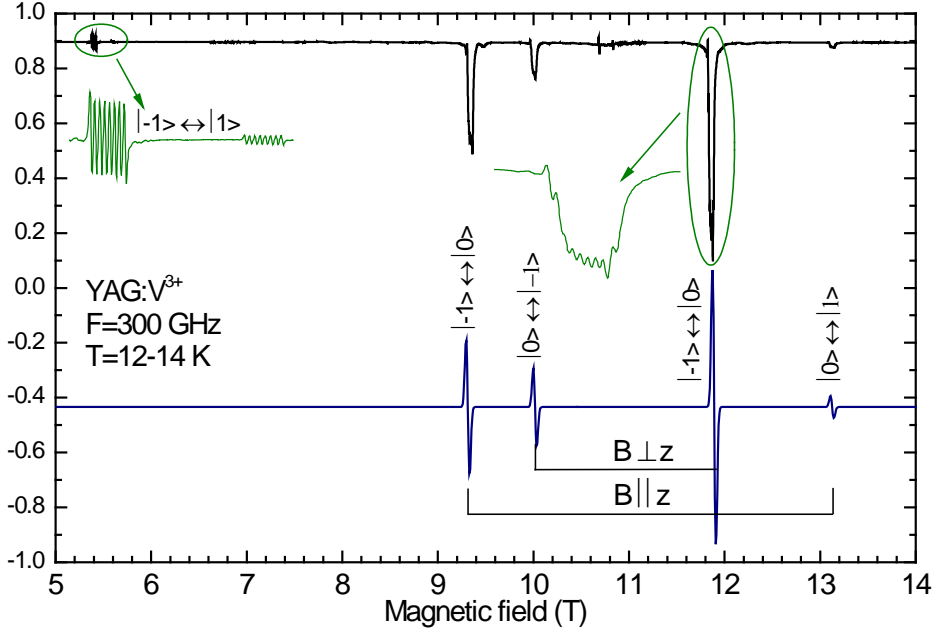

Fig. 5. Measured (upper curve) and calculated (bottom curve) tetrahedral $V^{3+}$ spectrum in YAG at the frequency 300 GHz. Only allowed transitions for only electron spin levels were taken in the calculated spectrum. Some of the measured spectral lines have an absorption-like shape due to saturation effects.

The spectrum taken at 300 GHz and experimental angular dependencies measured at 9.2 GHz were fitted by the calculated ones [34] using the following spin-Hamiltonian:

$$\hat{H}=\beta_e\left(g_\parallel \hat{S}_z B_z + g_\perp \hat{S}_x B_x + g_\perp \hat{S}_y B_y\right) + \frac{b_2^0}{3}\left(3\hat{S}_z^2 - S(S+1)\right), \tag{1}$$

where $g_\parallel$, $g_\perp$ are axial components of the $g$ tensor, $\hat{S}_x, \hat{S}_y$, and $\hat{S}_z$ are components of the electron spin $\hat{\mathbf{S}}$ ($S = 1$), and $b_2^0$ is the axial zero field splitting (ZFS) constant. The $z$ principal axis is parallel to the <001> crystal directions. Because the spectrum in Fig. 5 presents resonances for two orientations of the $V_T^{3+}$ center, all spin Hamiltonian parameters can be easily calculated. In particular, these parameters were calculated from fitting of the calculated spectrum to the measured one using numerical diagonalization [34] of the spin Hamiltonian (1). The following parameters were determined: $g_\parallel = 1.910 \pm 0.005$, $g_\perp = 1.950 \pm 0.005$, and $b_2^0 = -1.700 \pm 0.002$ cm$^{-1}$.

In the calculated spectrum, Boltzmann distribution of spin levels was taken into account that leads to different intensities of the $|-1\rangle \leftrightarrow |0\rangle$ and $|0\rangle \leftrightarrow |+1\rangle$ transitions allowing determination of zero-field splitting sign. It is negative.

Using the above determined spin Hamiltonian parameters the angular dependencies measured at 9.2 GHz were fitted also well (solid lines in Fig. 4). Finally, from the spectrum at 300 GHz, $^{51}$V HF splitting constants $A_{||}=67\pm0.5\times10^{-4}$ cm$^{-1}$; $A_{\perp}=68\pm0.5\times10^{-4}$ cm$^{-1}$ were calculated.

### 3.3.1. Crystal-field analysis of $V_T^{3+}$ spin Hamiltonian parameters

The determined spin-Hamiltonian parameters of the $V_T^{3+}$ center can be interpreted in the following way. $^3$F ground level of V$^{3+}$ free ion in a regular tetrahedron of the T$_d$ point symmetry will be split in the $^3$A$_2$ ground, $^3$T$_2$ first excited and $^3$T$_1$ second excited states, respectively, oppositely to the O$_h$ symmetry [32]. By experiencing a tetragonal distortion (D$_{2d}$ point group) which lifts the degeneracy of the $^3$T$_2 \rightarrow$ $^3$B$_2 + ^3$E and $^3$T$_1 \rightarrow ^3$A$_2 + ^3$E, the $^3$B$_1$ ground state will appear instead of the $^3$A$_2$ [33]. The corresponding energy level scheme is shown in Fig. 6.

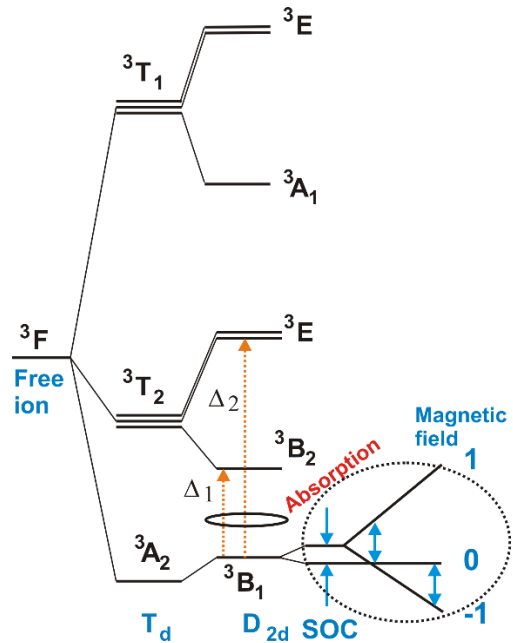

Fig. 6. Energy levels scheme of the 3d$^2$ ion in a tetrahedral symmetry crystal field. The energy separations do not correspond to the real ones, only a general tendency is hold. T$_d$ and D$_{2d}$ are the designations of the point symmetry groups. Dotted ellipse indicates a magnified part of the scheme as the splitting due to local crystal field is several orders stronger than the spin-orbital coupling (SOC) and magnetic field effects.

The set of the corresponding wave functions should thus be the following [35]:

$$B_1 : |n_0\rangle = \frac{\alpha}{\sqrt{2}}(|2\rangle - |-2\rangle) + \alpha'\phi_0,$$

$$B_2 : |n_1\rangle = \frac{\beta}{\sqrt{2}}(|2\rangle + |-2\rangle) + \beta'\phi_1,$$

$$E : |n_2\rangle = \gamma\left(\sqrt{\frac{5}{8}}|-1\rangle - \sqrt{\frac{3}{8}}|3\rangle\right) + \gamma'\phi_2,$$

$$|\tilde{n}_2\rangle = \gamma\left(\sqrt{\frac{5}{8}}|1\rangle - \sqrt{\frac{3}{8}}|-3\rangle\right) + \gamma'\tilde{\phi}_2,$$

(2)

where $\alpha$, $\beta$, $\gamma$, $\alpha'$, $\beta'$, $\gamma'$ are the weight coefficients of the corresponding $V_T^{3+}$ $d$ orbitals and $\phi_i, \tilde{\phi}_i$ are $2p$ ligand oxygen wave functions ($i = 1,2,3$). The explicit view of the ligand orbitals is unnecessary as they do not influence the expressions for the $g$ factors below according to considerations based on the linear combinations of atomic orbitals – molecular orbital (LCAO-MO) theory.

The $g$ factor and hyperfine tensor values then can be expressed in the way:

$$g_\| = g_e - \frac{8\lambda}{\Delta_1}\alpha^2\beta^2,$$

$$g_\perp = g_e - \frac{8\lambda}{\Delta_2}\alpha^2\gamma^2,$$

$$A_\| = A - P\left(g_e - g_\| - \frac{1}{112}(g_e - g_\perp)\right),$$

$$A_\perp = A - \frac{223}{224}P(g_e - g_\perp),$$

(3)

where $g_e = 2.0023$ is the free electron $g$ factor, $\Delta_{1,2}$ are the energy separations between the $|n_0\rangle$ ground and $|n_{1,2}\rangle$ excited states. $A$ is the Fermi contact term; $P = g_e g_n \beta_e \beta_n \langle r^{-3}\rangle$, $g_n(^{51}V) = 1.4711$ [36], $\beta_n$ are the $^{51}V$ nuclear $g$ factor and nuclear magneton, respectively; $r$ is a paramagnetic $3d$ electron-nucleus distance. The spin-orbital coupling (SOC) reduction coefficients are: $K_\| = \alpha^2\beta^2$ and $K_\perp = \alpha^2\gamma^2$. The $\Delta_{1,2}$ values were deduced from Fig. 1 in the way: $\Delta_1 = 7693$ cm$^{-1}$ (peak 5) and $\Delta_2 = 8697$ cm$^{-1}$ (peak 4) corresponding to the 1300 nm and 1150 nm strongly overlapped bands. From Eq. 3 the $K_\| = 0.42$ and $K_\perp = 0.27$. Hence the influence of the ligands is expected stronger in the plane perpendicular to the axial axis as $\gamma^2 \approx 0.64\beta^2$

. The charge distribution in the $V_T^{3+}$ ion is therefore not uniform that results from relatively large distortion of the tetrahedron (it is elongated along [100] directions by 5.7% [27]).

In the crystal-field approximation, the axial ZFS constant can be presented in the form: $b_2^0 = \frac{\lambda}{2}(g_\parallel - g_\perp)$ [32, 35]. Consequently, the calculated $b_2^0 = -4.2$ cm$^{-1}$ is about 2.5 times larger compared to the experimentally determined value of $-1.7$ cm$^{-1}$. The reduction of the spin-orbit coupling due to covalence in this case is expected to be approximately 40% which correlates perfectly well to the SOC reduction determined from the g factors. So the reduced SOC constant for the tetrahedrally coordinated $V^{3+}$ ion is $\lambda^* = 210 \times 0.4 = 84$ cm$^{-1}$. Remarkably, for the six-fold coordinated $V^{3+}$ in corundum, the $\lambda^*$ was determined as 38 cm$^{-1}$ [31]. Also the negative sign of $b_2^0$ constant determined experimentally agrees well with the relation $b_2^0 = \frac{\lambda}{2}(g_\parallel - g_\perp) < 0$ since $g_\parallel < g_\perp$.

$^{51}$V hyperfine parameters of the $V_T^{3+}$ center were analyzed (Eq. 3) considering $A_\parallel > 0$ and $A_\perp < 0$. Only in this way it was possible to obtain parameters which give a physical sense because the $P$ parameter should be positive as long as the free ion value, 450 MHz [36], is positive along with the positive nuclear g factor. The Fermi contact terms $A = 242$ MHz and $P = 447$ MHz were calculated. The $\chi = \frac{4\pi}{S}\left(\psi \left|\sum_i \delta(r_i) s_{zi}\right| \psi\right) = -\frac{3}{2}\frac{hca_0^3}{g_e g_n(^{51}V)\beta_e \beta_n} A = -3.62$ parameter characterizing the spin density at a nucleus was calculated. It obtained value which is in accordance with the values determined for the vanadium ions in other materials (see [36] and references therein).

### 3.4. EPR of $V^{4+}$ tetrahedral center

Annealing in air of the as-grown YAG:V sample leads to the disappearance of the $V_T^{3+}$ signals as well as $V_O^{3+}$ with the simultaneous growth of a new one instead (Fig. 3b). The new spectrum was assigned to the tetrahedral $V^{4+}$ center ($V_T^{4+}$, $3d^1$, $S = 1/2$) due to pronounced $^{51}$V HF structure and spin value 1/2. Only two spectral components due to three magnetically inequivalent positions of the center were visible at the magnetic field $\mathbf{B} \parallel [001]$. Such behavior of the $V^{3+}/V^{4+}$ EPR signals again confirms the correct suggestion about the origin of the signals and assignment of the peaks in Fig. 1 to the tetrahedral and octahedral $V^{3+}$. Annealing of as-grown crystal in $H_2$ atmosphere slightly reduces the $V_T^{3+}$ spectral intensity what is in good agreement with the corresponding absorption peaks in Fig. 1. Note that $V_O^{4+}$ EPR spectrum was not observed in our crystals. This does not mean that such centers are not created in YAG:V crystals after post-grown annealing in air, because the disappearance of $V_O^{3+}$ EPR spectrum directly suggests recharge of the $V^{3+}$ ions to the $V^{4+}$ charge state. The $V_O^{4+}$ ions can hardly be detected in our EPR measurements because the $d^1$ ion in an octahedral environment possesses the *g* factor which is close to zero because of the triply degenerate ground state unless the Jahn-Teller effect or low lattice symmetry remove the degeneracy [32]. But even in this case, the g factor value cannot be well predicted, and the experimental evidence is thus too complicated. In any case, the $V_O^{4+}$ signals were not observed in our crystals.

Angular dependencies of the $V_T^{4+}$ center resonance fields in the (010) plane are shown in Fig. 7. In this crystal plane, all three sets of resonances from three magnetically nonequivalent center positions are well resolved. They have been fitted by the calculated ones [34] using the following spin-Hamiltonian for the electron spin $S = 1/2$:

$$\hat{H}=\beta_e \left( g_\| \hat{S}_z B_z + g_\perp \hat{S}_x B_x \right) + A_\| \hat{S}_z B_z + A_\perp \hat{S}_x B_x, \tag{4}$$

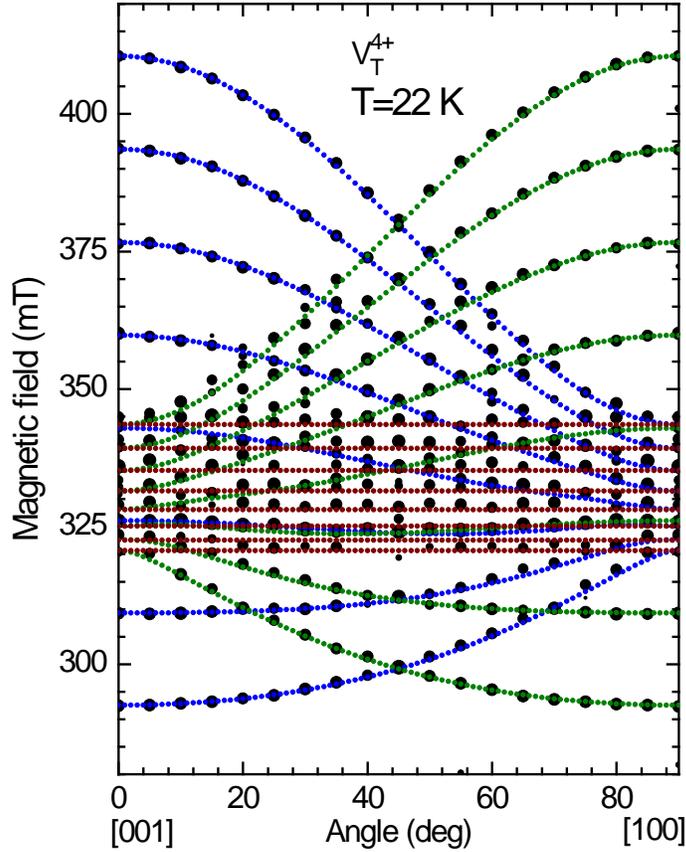

Fig. 7. Angular dependencies of the $V_T^{4+}$ resonance fields at crystal rotation in the (010) plane. Dots are the experimental data and solid dotted lines are the calculated data.

For the $V_T^{4+}$ center, similarly to $V_T^{3+}$ center, the main z-axis is parallel to the [100] crystal directions. The following spin-Hamiltonian parameters were determined: $g_\| = 1.881 \pm 0.003$, $g_\perp = 1.987 \pm 0.003$, $|A_\|| = (148 \pm 5) \times 10^{-4}$ cm$^{-1}$, $|A_\perp| = (30 \pm 1) \times 10^{-4}$ cm$^{-1}$.

### *3.4.1. Crystal-field analysis of $V_T^{4+}$ spin Hamiltonian parameters*

The analysis of the above determined parameters needs a similar approach to that reported previously for a Cu$^{2+}$ ion with the dominating $|d_{x^2-y^2}\rangle$ ground state [37]. The D$_{2d}$ distortion of the tetrahedron in YAG leads to the $|d_{x^2-y^2}\rangle$ orbital ground state since $g_\perp > g_\|$ [37]. The spin-Hamiltonian parameters after certain algebraic manipulation can be expressed in the way:

$$g_\| = g_e - \frac{8\lambda \xi^2 \chi^2}{\Delta_\|},$$

$$g_\perp = g_e - \frac{2\lambda \xi^2 \vartheta^2}{\Delta_\perp},$$

$$A_\| = A + P\left(-\frac{4\xi^2}{7} + g_\| - g_e + \frac{3}{7}(g_\perp - g_e)\right), \qquad (5)$$

$$A_\perp = A + P\left(\frac{2\xi^2}{7} + \frac{11}{14}(g_\perp - g_e)\right),$$

where $\xi$, $\chi$, $\vartheta$ are the constants accounting for the contributions of the $|d_{x^2-y^2}\rangle$ to the ground, $|d_{xy}\rangle$ to the first excited and $|d_{xz}\rangle$ to the second excited states, respectively; $\lambda(V^{4+}) = 250$ cm$^{-1}$ [38], $\Delta_\| = 16289$ cm$^{-1}$ (614 nm in Fig. 1, according to the analysis of the absorption spectra), $\Delta_\perp = 30309$ cm$^{-1}$ (330 nm in Fig. 1) are the energy separations between the ground and first and second excited states, respectively. Since $g_n(^{51}V) > 0$, P should be also positive. The $\Delta_\perp$ was deduced from the absorption spectra in Fig. 1 in the way, that the $\xi^2 \vartheta^2 = 1$ puts limitation to the $\Delta_\perp \leq 32680$ cm$^{-1}$ (306 nm) and since the band 1 survives thermal treatments without visible critical changes of its shape and position, the $V_T^{4+}$ second band should overlap with the absorption edge. So the 330 nm (Fig. 1) at its utmost low energy region seems to be a good choice taking into account reduction of the spin-orbital coupling due to covalent bonds with ligand creation.

Using thus defined splitting and determined spin-Hamiltonian parameters one can further obtain: $\xi^2\chi^2 = 0.93$ and $\xi^2\vartheta^2 = 0.93$. Since both the $\xi^2\chi^2$ and $\xi^2\vartheta^2$ products have the values very close to 1, the $\xi^2 \approx \chi^2 \approx \vartheta^2 = \sqrt{0.93} = 0.96$ was assumed. Only $A_\parallel, A_\perp < 0$ give a physical sense to the hyperfine parameters, whereas other possibilities resulted in $P$ significantly larger and/or negative (it should be positive in accordance with the free $V^{4+}$ ion value $P$ = 516 MHz [36] and positive nuclear $g$ factor). Besides, consulting to the data reported in many other hosts (see [36] and references therein) the $A_\parallel$ and $A_\perp$ are both negative in most cases for the vanadium ions. Then from Eq. 5, $A = -189$ MHz, $P = 376$ MHz. Comparing the determined $P$ with the free ion value along with significantly dominating vanadium $d$ orbital in the ground and excited states' wave functions ($\xi^2$ value is very close to 1) one can conclude that the electron density is predominately concentrated on the ion with a very small distribution over ligand oxygens in a tetrahedron. The calculated parameter characterizing spin density at a nucleus $\chi = -\frac{3}{2}\frac{hca_0^3}{g_e g_n(^{51}V)\beta_e\beta_n}A = 2.83$ is comparable with the values determined for the $V^{4+}$ ion in other materials (see [36] and references therein).

## 4. Conclusions

Optical absorption and electron paramagnetic resonance measurements have shown that the vanadium ions charge state in the as-grown YAG crystals is predominantly 3+ at both the octahedral and tetrahedral aluminum lattice sites. Post-grown annealing of YAG:V crystals in air causes the $V^{3+}$ to $V^{4+}$ charge transformation. This was directly confirmed by the measurements of EPR spectra of these ions. Annealing in a reduction ($H_2$) atmosphere only slightly reduces the $V^{3+}$ absorption bands amplitude and corresponding EPR signals probably due to lowering of the actual charge state of only a small part of $V^{3+}$ ions.

Using the high-frequency (up to 300 GHz) EPR measurements the spin-Hamiltonian parameters including the $^{51}V$ hyperfine and zero field splitting constants of the $V_T^{3+}$ center were determined. Analyzing them in the frameworks of crystal field and LCAO-MO theory in cooperation with optical absorption and exploiting $D_{2d}$ distortion of the regular tetrahedron the charge distribution over the $V^{3+}$ and ligand oxygens was found to be strongly anisotropic. In the plane perpendicular to the axial axis, the ligands are playing more significant role than the central vanadium ion as compared to the covalency along the axial axis. The determination of precise contributions of the ligands was hardly possible basing on our experimental data. However, resulting reduction of the spin orbit coupling is estimated to be around 40%.

Both the g and $^{51}$V HF tensors were determined for the $V_T^{4+}$ ions in YAG. These ions exhibit only very weak covalent bonding with ligands, about 4%.

The present study shows that while the optical absorption peaks from vanadium ions of different valence states at different lattice sites are difficult to be interpreted due to the overlapping of peaks, the EPR spectra of these vanadium ions are well separated one from other. Consequently, the proper assignment of the optical absorption peaks in YAG:V crystals was made by taking into account EPR data including the g tensor parameters analysis.

## Acknowledgements

The authors gratefully acknowledge the financial support of the Czech Science Foundation (project No. 17-09933S), the Ministry of Education, Youth and Sports of Czech Republic under project LO1409 and CZ.02.1.01/0.0/0.0/16_013/0001406. We also thank J. van Slageren (Stuttgart) for providing access to the high-frequency EPR spectrometer. OL is grateful to the Baden-Württemberg Stiftung for the financial support by the Eliteprogramme for Postdocs.